\documentclass[conference,9.5pt]{IEEEtran}
\IEEEoverridecommandlockouts

\usepackage{cite}
\usepackage{amsmath,amssymb,amsfonts}
\usepackage{algorithmicx}
\usepackage{graphicx}
\usepackage{textcomp}
\usepackage{xcolor}
\usepackage{multirow}
\usepackage{pifont}
\usepackage{graphicx}
\usepackage{algorithm}
\usepackage{algpseudocode}

\newcommand{\tabincell}[2]{\begin{tabular}{@{}#1@{}}#2\end{tabular}}
\def\BibTeX{{\rm B\kern-.05em{\sc i\kern-.025em b}\kern-.08em
    T\kern-.1667em\lower.7ex\hbox{E}\kern-.125emX}}
\begin{document}

\title{Phone-purity Guided Discrete Tokens for Dysarthric Speech Recognition}

\author{
\IEEEauthorblockN{Huimeng Wang$^1$, Xurong Xie$^{2\ast}$, Mengzhe Geng$^3$, Shujie Hu$^1$, Haoning Xu$^1$, Youjun Chen$^1$, \\Zhaoqing Li$^1$, Jiajun Deng$^1$, Xunying Liu$^{1\ast}$\thanks{*Corresponding author}}
\IEEEauthorblockA{
\textit{$^{1}$The Chinese University of Hong Kong, Hong Kong SAR, China}\\
\textit{$^{2}$Institute of Software, Chinese Academy of Sciences, Beijing, China}\\
\textit{$^3$National Research Council of Canada, Canada}}
}
\maketitle
\begin{abstract}
Discrete tokens extracted provide efficient and domain adaptable speech features.
Their application to disordered speech that exhibits articulation imprecision and large mismatch against normal voice remains unexplored.
To improve their phonetic discrimination that is weakened during unsupervised K-means or vector quantization of continuous features, this paper proposes novel phone-purity guided (PPG) discrete tokens for dysarthric speech recognition. 
Phonetic label supervision is used to regularize maximum likelihood and reconstruction error costs used in standard K-means and VAE-VQ based discrete token extraction.
Experiments conducted on the UASpeech corpus suggest that the proposed PPG discrete token features extracted from HuBERT consistently outperform hybrid TDNN and End-to-End (E2E) Conformer systems using non-PPG  based K-means or VAE-VQ tokens across varying codebook sizes by statistically significant word error rate (WER) reductions up to 0.99\% and 1.77\% absolute (3.21\% and 4.82\% relative) respectively on the UASpeech test set of 16 dysarthric speakers.
The lowest WER of 23.25\% was obtained by combining systems using different token features.
Consistent improvements on the phone purity metric were also achieved.
T-SNE visualization further demonstrates sharper decision boundaries were produced between K-means/VAE-VQ clusters after introducing phone-purity guidance.

\end{abstract}

\begin{IEEEkeywords}
Speech Disorders, Speech Recognition, Discrete Tokens, Speech Foundation Models 
\end{IEEEkeywords}

\vspace{-2mm}
\section{Introduction}
\label{sec:intro}
Despite the rapid progress of automatic speech recognition (ASR) technologies targeting normal speech, accurate recognition of pathological voice, for example, dysarthric speech, remains a highly challenging task \cite{inproceedings,xiong20source,liu21recent,mengzhetaslp,wang23y_interspeech,shujie23taslp,wang2024enhancing} to date due to: a) the scarcity of such data; b) its large mismatch against the normal speech; c) large speaker level diversity.
The physical disabilities and mobility issues associated with impaired speakers exacerbate the challenges of collecting large quantities of dysarthric speech for ASR system development.
To this end, a recent series of researches have focused on transferring knowledge from ASR systems trained on normal speech to address dysarthric speech recognition tasks\cite{inproceedings,xiong2019phonetic,xiong20source,liu21recent,geng2020investigation,wang23y_interspeech,mengzhetaslp,shujie23taslp,wang2024enhancing,hu2023exploring,wang2023benefits,jin21_interspeech,jin2024rl}.

In recent years, self-supervised learning (SSL) based speech foundation models \cite{baevski2019vq,baevski2020wav2vec,hsu2021hubert,chen2022wavlm} pre-trained on massive unlabeled data have emerged as a powerful paradigm for multiple downstream speech processing tasks. 
Neural speech representations from these models are highly adaptable across different task domains.
In particular, discrete token features, which provide compact speech representations, have been successfully applied to a wide variety of downstream tasks including, but not limited to speech recognition \cite{chang23b_interspeech,chang2024exploring,yang2024towards,sukhadia24_interspeech}, text-to-speech synthesis\cite{du22b_interspeech,unicats,yang2024towards}, and speech resynthesis \cite{nguyen23_interspeech}.
Recent study \cite{sukhadia24_interspeech} on children's speech recognition shows the potential of discrete token features for low-resource speech tasks.

In contrast, the application of discrete tokens to pathological speech domains, e.g. dysarthric speech recognition, remains under study.
Recent studies\cite{chang2024exploring,chang23b_interspeech,yang2024towards,sukhadia24_interspeech} report significant performance degradation when comparing discrete token based ASR systems against those using comparable continuous, un-quantized SSL speech representations extracted from the same foundation models.
One important reason behind such performance degradation over continuous features is that their lack of supervision information during unsupervised feature quantization using, e.g. K-means and Gumbel-softmax based vector quantization (VQ).
This drawback leads to weakened discrimination over, e.g. phonetic units.
In particular, this issue is further exasperated by the widely observed articulation imprecision \cite{KENT2000273} in dysarthric speech data.

To this end, this paper proposes novel phone-purity guided discrete token extraction approaches for dysarthric speech recognition.
Frame-level phonetic label supervision is used to regularize the maximum likelihood and reconstruction error costs used in standard K-means and VAE-VQ based discrete token extraction process.

Experiments conducted on the largest publicly available and most widely used UASpeech \cite{uaspeech2008} suggest that the proposed PPG discrete token features extracted from HuBERT consistently outperform hybrid TDNN and End-to-End (E2E) Conformer systems using non-PPG  based K-means or VAE-VQ tokens across varying codebook sizes by statistically significant word error rate (WER) reductions up to 0.99\% and 1.77\% absolute (3.21\% and 4.82\% relative) respectively on the UASpeech test set of 16 dysarthric speakers.
The lowest WER of 23.25\% was obtained by combining systems using different token features.
Consistent improvements on the phone purity metric were also achieved.
T-SNE visualization further demonstrates sharper decision boundaries were produced between K-means and VAE-VQ clusters after introducing phone-purity guidance during quantization.

The main contributions of this paper are summarized below:

\noindent \textbf{1)} To the best of our knowledge, this paper presents the first study on the use of discrete token features from speech foundation models \cite{baevski2019vq,baevski2020wav2vec,hsu2021hubert,chen2022wavlm,hu2023exploring} for dysarthric speech recognition tasks. 
In contrast, limited prior researches mainly focus on utilizing continuous SSL speech representations \cite{hernandez22_interspeech,zrvae,AVSSL,benefits,shujie23taslp} to improve dysarthric speech recognition performance, while discrete tokens have not been studied for such data.




\noindent \textbf{2)} The discrete tokens with phone-purity guidance consistently outperform the TDNN and Conformer systems constructed using non-phone-purity guided K-means or VAE-VQ token features across different codebook sizes by statistically significant WER reductions up to 0.99\% and 1.77\% absolute (3.21\% and 4.82\% relative) respectively on UASpeech task.
After system combination, the best-performing system featuring different token features produces the lowest WER 23.25\%.

\noindent \textbf{3)} These WER reductions are strongly correlated with the improvements obtained on the phone purity metric. 
T-SNE visualization further demonstrates sharper decision boundaries were produced between K-means and VAE-VQ clusters after introducing phone-purity guidance during quantization.

\begin{figure}
    \centering
    \includegraphics[width=1.0\linewidth]{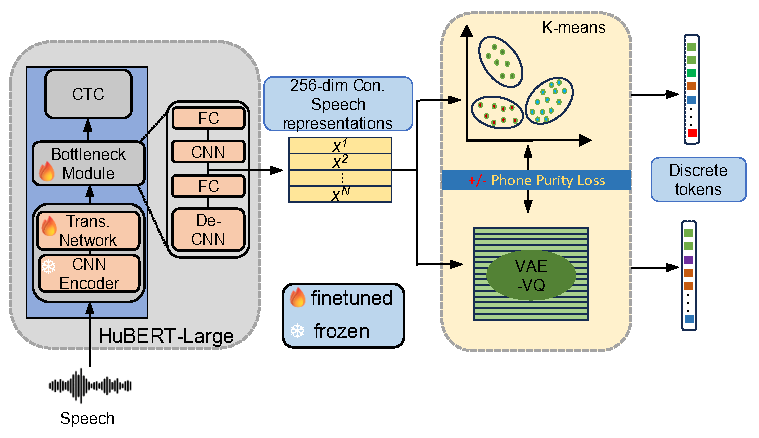}
    \vspace{-8mm}
    \caption{Illustration of discrete token extraction from fine-tuned HuBERT models. The 256-dim compact continuous speech representations are quantized into discrete tokens via either K-means or VAE-VQ without(\textcolor{red}{-})/with(\textcolor{red}{+}) phone purity loss regularization.}
    \label{fig:1}
    \vspace{-6mm}
\end{figure}

\section{Foundation Models and Discrete Tokens}
\subsection{HuBERT}
\noindent\textbf{Model architecture:} HuBERT model consists of three components, including 1) a CNN feature encoder to encode raw speech $\mathcal{O}$ into continuous representations $z_t \in \mathcal{Z}$ with a $20$ms stride and a $25$ms receptive field, 
2) a transformer network to produce context representations $c_t \in C$ over the entire sequence of randomly masked feature encoder outputs followed by a projection layer, and 3) a K-means quantization module to generate speech units $q_t \in \mathcal{Q}$ as pseudo-labels for self-supervised training learning or pre-training.

\noindent\textbf{Model pre-training:} HuBERT pre-training process alternates between two steps: 1) an offline clustering step on initially selected MFCC features or subsequently selected embedding features from the intermediate layer of the transformer network to create frame-level pseudo-labels and 2) a masked prediction step to optimize a BERT-like loss based on pseudo-labels.
By such an iterative refinement, HuBERT learns a unified language and acoustic model from raw speech directly\cite{hsu2021hubert}.

\noindent\textbf{Model fine-tuning:} During fine-tuning, a randomly initialized Softmax layer replaces the projection layer. 
The model parameters are fine-tuned on labeled speech data using Connectionist Temporal Classification (CTC) loss $\mathcal{L}_{CTC}$ with the CNN-based feature encoder frozen.

\subsection{Discrete token extraction}
As shown in Fig.~\ref{fig:1}, the discrete token feature extraction can be divided into two sequential stages: 1) continuous neural speech representation extraction and 2) further quantizing such neural features into discrete tokens.

\noindent\textbf{Continuous neural speech representations} are extracted from raw speech via a cross-domain fine-tuned HuBERT model.
During fine-tuning, a bottleneck module is inserted between the Transformer network and the CTC module to produce more compact neural speech representations.
The bottleneck module consists of a stack of four interleaving convolutional and feedforward layers: 1) a 1D transposed de-convolution CNN layer changing the stride length from 20ms to 10ms, which allows frame synchronization with the back-end ASR systems; 2) a fully connected (FC) block, consisting of a linear layer, rectified linear unit (ReLU) activation, and dropout module, to change the embedding dimension from 1024 to 256; 3) a second CNN layer to revert the stride to 20ms; and 4) a second FC block to restore the dimensionality to 1024.
The final HuBERT speech representations are extracted from the first FC layer outputs.

\noindent\textbf{Discrete tokens} are quantized from continuous speech representations via codebook generated by either K-means or VAE-VQ.
More specifically, each frame of an utterance is replaced by selecting the corresponding codeword from the codebook.

\section{phone-purity guided discrete tokens}
In this section, we present phone-purity guided K-means and VAE-VQ discrete token extraction methods.





\subsection{Phone-purity guided K-means quantization}\label{kmeans}
Given a set of un-quantized frame level continuous features extracted via HuBERT (or more generally other speech foundation models, e.g. WavLM \cite{chen2022wavlm} or Whisper \cite{pmlr-v202-radford23a}), ${\cal X} = \left\{\boldsymbol{x}_1,\boldsymbol{x}_2,..,\boldsymbol{x}_T\right\}$ of $T$ frames in total, and their reference phonetic labels ${\cal Y} = \left\{y_1,y_2,...,y_T \right\}$ derived using  force alignment, the proposed phone-purity guided K-means quantization alternates between a standard K-means cluster assignment step and a modified phone-purity regularized cluster means update step until convergence. 
At the $l^{th}$ clustering iteration, let
${\cal C}^{(l)} = \left\{\boldsymbol{c}_{1}^{(l)}, \boldsymbol{c}_{2}^{(l)}, ..., \boldsymbol{c}_{K}^{(l)} \right\}$ and 
$\Gamma^{(l)} = \left \{\gamma_{1}^{(l)}, \gamma_{2}^{(l)}, ..., \gamma_{T}^{(l)} \right\}$ denote the current $K$ cluster specific mean vectors,  
and the $T$ frame level quantized cluster indices, respectively.  

\noindent\textbf{1) Cluster assignment step} infers the most probable frame level quantized labels as follows
\begin{equation}
\setlength\abovedisplayskip{1pt}
\setlength\belowdisplayskip{1pt}
\gamma_{t}^{(l)} ~=~  \arg \min_{1 \leq k \leq K} \left\{ \left\|\boldsymbol{x}_{t} - \boldsymbol{c}_{k}^{(l)} \right\|^{2} \right\}
\label{eqn:KMeans_Assign}
\end{equation}


\noindent\textbf{2) Cluster means update step} re-estimates the cluster centroids (i.e. quantized vector code-book entries) at the $(l+1)^{th}$ iteration by incorporating an additional phone-purity regularization term as below
\newpage
\begin{eqnarray}
\setlength\abovedisplayskip{0pt}
\setlength\belowdisplayskip{0pt}
 \!\!\! \boldsymbol{c}_{k}^{(l+1)} \!\!\! &   \!\!=  \!\!&  \!\! \arg \min  \! \left\{  \!
\sum_{t=1,  \gamma_{t}^{(l)} = k}^{T}  
\!\!\!  \!\!\! \! \left\| \boldsymbol{x}_{t}  \!\ - \! \boldsymbol{c}_{k}^{(l+1)} \right\|^{2} 
\! +  \! \lambda  \left\| \boldsymbol{p}_{k}^{(l)} \!-\! \boldsymbol{c}_{k}^{(l+1)} \right\|^{2}
 \! \right\} \nonumber\\
 &  \!\!\! =  \!\!\! & \frac{\sum_{t=1,\gamma_{t}^{(l)} = k}^{T} \boldsymbol{x}_{t} + \lambda \boldsymbol{p}_{k}^{(l)} }{ \sum_{t=1, \gamma_{t}^{(l)} = k}^{T} 1 + \lambda}
\label{eqn:KMeans_Update}
\end{eqnarray}
where the phone-purity regularization in the second term is obtained by computing the {\bf purest cluster centroid}, $\boldsymbol{p}_{k}^{(l)}$, as the mean average over all the frame level un-quantized features that are currently assigned to the $k^{th}$ cluster at the $l^{th}$ iteration, and share the top-1 most frequent reference phonetic label.    

The above iterative phone-purity guided K-means clustering converges when the total 
cluster level centroids differences 
between neighboring clustering iterations is sufficiently small,
$\sum_{k=1}^{K} \left\| \boldsymbol{c}_{k}^{(l)} - \boldsymbol{c}_{k}^{(l+1)} \right\| ^{2}
\leq \epsilon$, 
where the residual error is empirically set as $\epsilon = 1e-5$ in this paper. In practice, more stable convergence can be obtained by initializing the centroids using the K-means++ method\cite{kmeans++} before the above starts.

\vspace{-1mm}
\subsection{Phone-purity guided VAE-VQ quantization}
The standard reconstruction loss used in VAE-VQ training is augmented using a phone-purity regularization term as
\begin{equation}
\setlength\abovedisplayskip{1pt}
\setlength\belowdisplayskip{1pt}
    {\cal L}_{\rm VAE-VQ} ~=~ {\cal L}_{\rm MSE} + \alpha {\cal L}_{\rm Pur} 
    \label{eqn:loss_vae_vq}
\end{equation}
In the above, the first MSE based VAE-VQ {\bf reconstruction loss} at the $l^{th}$ SGD batch update using $T_l$ frames of data is  
\begin{equation}
\setlength\abovedisplayskip{1pt}
\setlength\belowdisplayskip{0pt}
    \mathcal{L}_{\rm MSE} ~=~ \frac{1}{T_l}\sum_{t=1}^{T_l} \left\|\boldsymbol{x}_t-\boldsymbol{c}^{(l)}_{\gamma_t^{(l)}}\right\|^{2}
    \label{eqn:loss_vae_vq_mse}
\end{equation}
while the frame level VQ code-book index, $\gamma_t^{(l)}$, is inferred using Equation~(\ref{eqn:KMeans_Assign}).

The {\bf phone-purity regularization} term, ${\cal L}_{\rm Pur}$, in the above Equation (\ref{eqn:loss_vae_vq}) is the entropy loss computed using the VAE-VQ quantized features against the reference phonetic labels.
\begin{equation}
\setlength\abovedisplayskip{1pt}
\setlength\belowdisplayskip{0pt}
    \mathcal{L}_{\rm Pur} = - \frac{1}{T_l}\sum_{t=1}^{T_l} 
    \sum_{j=1}^{N_p} P\left(y_j \mid \boldsymbol{c}^{(l)}_{\gamma_t^{(l)}}\right)\log P\left(y_j \mid \boldsymbol{c}^{(l)}_{\gamma_t^{(l)}}\right)
    \label{eqn:loss_vae_vq_pur}
\end{equation}
where $N_p$ denotes the number of reference phonetic labels. The phonetic label posterior probabilities conditioned on the quantized features are computed as
\begin{equation}
\setlength\abovedisplayskip{1pt}
\setlength\belowdisplayskip{1pt}
    P\left(y_j \mid \boldsymbol{c}^{(l)}_{\gamma_t^{(l)}}\right) ~=~ \dfrac{\mathcal{N} \left(\boldsymbol{c}^{(l)}_{\gamma_t^{(l)}}\mid \boldsymbol{\mu}_j, \boldsymbol{\Sigma}_j \right) }{\Sigma_{i=1}^{N_p} \mathcal{N} \left( \boldsymbol{c}^{(l)}_{\gamma_t^{(l)}}\mid \boldsymbol{\mu}_i, \boldsymbol{\Sigma}_i \right)}
    \label{eqn:phn_post}
\end{equation}
where $\mathcal{N}\left(\cdot \mid \boldsymbol{\mu}_j, \boldsymbol{\Sigma}_j \right)$ is the Gaussian PDF of the $j^{th}$ phonetic label class, parameterized by its mean, $\boldsymbol{\mu}_j$ and covariance, $\boldsymbol{\Sigma}_j$.


The above phone-purity guided VAE-VQ loss is 
minimized using SGD~\cite{ruder2016overview}. 
The phone-purity regularization scaling in Equation (\ref{eqn:loss_vae_vq}) is empirically set as $\alpha = 1.2$ and $\alpha = 1.05$ for $K=100$ and $K=500$ in this paper respectively.

\section{Experiments}
\begin{table}[!h]
    \caption{Performance of fine-tuned HuBERT (HuB.) based ASR,  TDNN and Conformer (CONF.) based systems constructed without or with(+) domain fine-tuned HuBERT features (HuB. Feat.) and system combination (Sys. Comb.) on the dysarthric UASpeech test set.
    ``SSL" denotes the continuous HuBERT speech representations.
    ``Ppg. K-means" and "Ppg. VAE-VQ" are the abbreviations of phone-purity guided K-means and phone-purity guided VAE-VQ.
    ``Code. Size" denotes the codebook size.
    ``VL", ``L", ``M" and ``H" denote intelligibility subgroups (Very Low/Low/Mild/High).
    ``$\dag$", ``$\diamond$" and ``$\star$" denote statistically significant (MAPSSWE\cite{mapsswe}, $\alpha=0.05$) improvement is obtained over the corresponding baseline systems (Sys. 4,8,12,18,22,26 for $\dag$, Sys. 8 for $\diamond$ and Sys. 22 for $\star$).
    ``+" represents frame-level joint decoding for Sys. 30-31 and score interpolation for Sys. 32-36 respectively.
    ``X$\rightarrow$Y denotes two pass rescoring \cite{cui22_interspeech} the $N$-best ($N=30$) outputs of system X by system Y.
    }
    \vspace{-2mm}
    \label{tab:results}
    \centering
     \renewcommand\arraystretch{1.1}
    \renewcommand\tabcolsep{1.0pt}
\resizebox{0.99\linewidth}{!}{
    \begin{tabular}{c|c|c|c|c|c|c|c|c|cccc|c}
    \hline\hline
    \multirow{4}{*}{Sys.}  &
    \multirow{4}{*}{Model} &
    \multirow{4}{*}{FBK} &
    \multicolumn{6}{c|}{HuB. Feat.} &
    \multicolumn{5}{c}{Word Error Rate \%} \\
    \cline{4-14}
    & & 
     & 
    \multirow{3}{*}{SSL} & 
    \multicolumn{5}{c|}{Discrete Token} &
    \multicolumn{4}{c|}{Intell. Subgroup} & \multirow{3}{*}{All} \\
    \cline{5-13}
    & & & & \multirow{1}{*}{K-means} & \tabincell{c}{Ppg.\\ K-means} & \multirow{1}{*}{VAE-VQ} & \tabincell{c}{Ppg. \\VAE-VQ} & \tabincell{c}{Code.\\ Size}& VL & L & M & H & \\
    \hline\hline
    1 & HuB. \cite{shujie23taslp} &\ding{55}&\ding{55} & \ding{55} &\ding{55}&\ding{55}&\ding{55}& - & 59.47 & 33.62 & 22.22 & 6.34& 27.71\\
    \cline{1-14}
    2 & \multirow{1}{*}{TDNN} &\ding{51} & \ding{55} & \ding{55} & \ding{55} & \ding{55} & \ding{55} & -  &  62.53 & 31.92 & 23.12 & 13.67 & 30.56 \\
    \cline{1-14}
    3 & \multirow{13}{*}{\tabincell{c}{TDNN+\\HuB.\\Feat.}} & \ding{55} & \ding{51} &\ding{55} & \ding{55} & \ding{55} & \ding{55} & - &  56.61  & 32.23  & 23.08  & 12.38  & 28.95  \\
    \cline{1-1}\cline{3-14}
    4 & &\ding{55} & \ding{55} & \ding{51} & \ding{55} & \ding{55} & \ding{55} & \multirow{4}{*}{100} &  60.82  & 35.50  & 25.69  & 13.55  & 31.57  \\
    5 & &\ding{55} & \ding{55} & \ding{55} & \ding{51} & \ding{55} & \ding{55} & &  60.50  & 35.48  & 25.24$^\dag$  & 12.71$^\dag$  & 30.86$^\dag$  \\
    6 & & \ding{55} & \ding{55} &\ding{55} & \ding{55} & \ding{51} & \ding{55} & &  60.18$^\dag$  & 34.76$^\dag$  & 24.39$^\dag$  & 12.39$^\dag$  & 30.59$^\dag$  \\
    7 & & \ding{55} & \ding{55} &\ding{55} & \ding{55} & \ding{55} & \ding{51} & &  59.89$^\dag$  & 34.15$^\dag$  & 24.31$^\dag$  & 12.12$^\dag$  & 30.29$^\dag$  \\
    \cline{1-1}\cline{3-14}
    8 & &\ding{55} & \ding{55} & \ding{51} & \ding{55} & \ding{55} & \ding{55} & \multirow{8}{*}{500} &  60.25  & 34.96  & 24.75  & 12.65  & 30.84  \\
    9 & & \ding{55} & \ding{55} &\ding{55} & \ding{51} & \ding{55} & \ding{55} & &  60.04  & 33.87$^\dag$  & 22.98$^\dag$  & 11.71$^\dag$  & 29.85$^\dag$  \\
    10 & &\ding{55} & \ding{55} & \ding{55} & \ding{55} & \ding{51} & \ding{55} & &  60.07  & 35.64  & 24.45  & 11.83$^\dag$  & 30.64  \\
    11 & & \ding{55} & \ding{55} &\ding{55} & \ding{55} & \ding{55} & \ding{51} & &  59.79$^\dag$  & 34.50  & 24.08  & 11.91$^\dag$   & 30.24$^\dag$  \\
    \cline{1-1}\cline{3-8}\cline{10-14}
    12 & & \ding{55} & \ding{51} &\ding{51} & \ding{55} & \ding{55} & \ding{55} & &  57.72  & 31.81  & 22.94  & 12.20  & 28.99  \\
    13 & & \ding{55} & \ding{51} &\ding{55} & \ding{51} & \ding{55} & \ding{55} & &  57.25  & 32.40  & 22.33$^\dag$  & 11.56$^\dag$  & 28.71$^\dag$  \\
    14 & & \ding{55} & \ding{51} &\ding{55} & \ding{55} & \ding{51} & \ding{55} & &  57.22  & 32.71  & 23.59  & 13.12  & 29.55  \\
    15 & & \ding{55} & \ding{51} &\ding{55} & \ding{55} & \ding{55} & \ding{51} & &  56.92$^\dag$ & 32.58 & 23.39 & 12.24 & 29.11 \\
    \hline\hline
    16 & \multirow{1}{*}{\tabincell{c}{CONF.}} & \ding{51} & \ding{55} & \ding{55} & \ding{55} & \ding{55} & \ding{55} & - & 66.06 & 48.4  & 46.35 & 41.6  & 49.45 \\
    \cline{1-14}
    17 & \multirow{12}{*}{\tabincell{c}{CONF.+\\HuB.\\ Feat.}} & \ding{55} & \ding{51} & \ding{55} & \ding{55} & \ding{55} & \ding{55} & - &  61.58  & 38.47  & 29.31  & 12.33  & 32.80  \\
    \cline{1-1}\cline{3-14}
    18 &  &\ding{55} & \ding{55} & \ding{51} & \ding{55} & \ding{55} & \ding{55} & \multirow{4}{*}{100} &  63.90  & 41.29  & 33.82  & 17.82  & 36.74  \\
    19 & & \ding{55} & \ding{55} &\ding{55} & \ding{51} & \ding{55} & \ding{55} & &  63.19$^\dag$  & 40.39$^\dag$  & 32.35$^\dag$  & 14.56$^\dag$  & 34.97$^\dag$  \\
    20 & & \ding{55} & \ding{55} &\ding{55} & \ding{55} & \ding{51} & \ding{55} & &  63.54  & 40.76$^\dag$  & 31.47$^\dag$  & 14.43$^\dag$  & 34.93$^\dag$  \\
    21 & & \ding{55} & \ding{55} &\ding{55} & \ding{55} & \ding{55} & \ding{51} & &  62.97$^\dag$  & 40.39$^\dag$  & 31.50$^\dag$  & 13.42$^\dag$  & 34.38$^\dag$  \\
    \cline{1-1}\cline{3-14}
    22 & &\ding{55} & \ding{55} & \ding{51} & \ding{55} & \ding{55} & \ding{55} & \multirow{8}{*}{500} &  63.42  & 39.92  & 31.56  & 13.40  & 34.36  \\
    23 & & \ding{55} & \ding{55} &\ding{55} & \ding{51} & \ding{55} & \ding{55} & &  62.97$^\dag$   & 39.92  & 31.50  & 12.94$^\dag$   & 34.10$^\dag$   \\
    24 & &  \ding{55} & \ding{55} &\ding{55} & \ding{55} & \ding{51} & \ding{55} & & 63.19  & 40.72  & 32.07  & 13.71  & 34.72  \\
    25 & & \ding{55} & \ding{55} &\ding{55} & \ding{55} & \ding{55} & \ding{51} & &  63.20  & 40.37  & 31.78  & 13.32  & 34.44  \\
    \cline{1-1}\cline{3-8}\cline{10-14}
    26 & &\ding{55} & \ding{51} & \ding{51} & \ding{55} & \ding{55} & \ding{55} & &  61.83  & 39.14  & 30.76  & 12.56  & 33.38  \\
    27 & &\ding{55} & \ding{51} & \ding{55} & \ding{51} & \ding{55} & \ding{55} & &  61.51  & 38.93  & 30.31$^\dag$  & 11.89$^\dag$  & 32.94$^\dag$  \\
    28 & & \ding{55} & \ding{51} &\ding{55} & \ding{55} & \ding{51} & \ding{55} & &  62.45  & 40.08  & 31.21  & 12.95  & 33.97  \\
    29 & &\ding{55} & \ding{51} & \ding{55} & \ding{55} & \ding{55} & \ding{51} & &  62.47  & 39.98  & 30.86  & 12.43  & 33.71 \\
    \hline\hline
    30 & \multirow{7}{*}{\tabincell{c}{Sys.\\Comb.}}  & \multicolumn{7}{c|}{Sys. 2 + 12 + 14} & 53.89$^\diamond$  & 26.26$^\diamond$  & 17.18$^\diamond$ & 8.84$^\diamond$  & 24.49$^\diamond$  \\
    31 & & \multicolumn{7}{c|}{Sys. 2 + 13 + 15} & 53.48$^\diamond$  & 26.30$^\diamond$  & 17.14$^\diamond$  & 8.83$^\diamond$  & 24.41$^\diamond$  \\
    \cline{1-1}\cline{3-14}
    32 & & \multicolumn{7}{c|}{Sys. (26 $\rightarrow$ 16) + 26 + (26 $\rightarrow$ 28)} & 62.05$^\star$& 39.11$^\star$& 30.57$^\star$& 12.40$^\star$ & 33.33$^\star$ \\
    33 & & \multicolumn{7}{c|}{Sys. (27 $\rightarrow$ 16) + 27 + (27 $\rightarrow$ 29)} & 61.71$^\star$& 38.98$^\star$ & 29.67$^\star$ & 11.56$^\star$& 32.77$^\star$ \\
    \cline{1-1} \cline{3-14}
    34 & & \multicolumn{7}{c|}{Sys. 30 + 31} & 53.65$^\diamond$ & 26.23$^\diamond$ & 16.96$^\diamond$ & 8.74$^\diamond$ & 24.36$^\diamond$ \\
    35 & & \multicolumn{7}{c|}{Sys. 32 + 33} & 61.82$^\star$ & 38.95$^\star$ & 29.65$^\star$& 11.59$^\star$ & 32.79$^\star$ \\
    \cline{1-1}\cline{3-14}
    36 & & \multicolumn{7}{c|}{Sys. 34 + 35}  & 54.14$^{\diamond\star}$& 25.68$^{\diamond\star}$& 16.33$^{\diamond\star}$& 5.92$^{\diamond\star}$ & 23.25$^{\diamond\star}$\\
    \hline\hline
    \end{tabular}
}
\vspace{-7mm}
\end{table}

\begin{figure*}[!th]
	\centering
	\begin{minipage}[t]{0.19\linewidth}
		\includegraphics[width=1.0\linewidth]{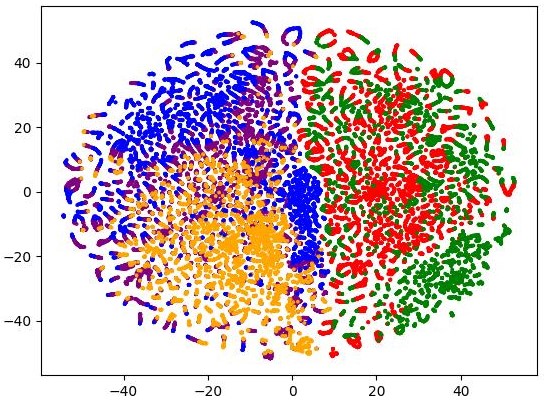}
	\end{minipage}
	\begin{minipage}[t]{0.19\linewidth}
		\includegraphics[width=1.0\linewidth]{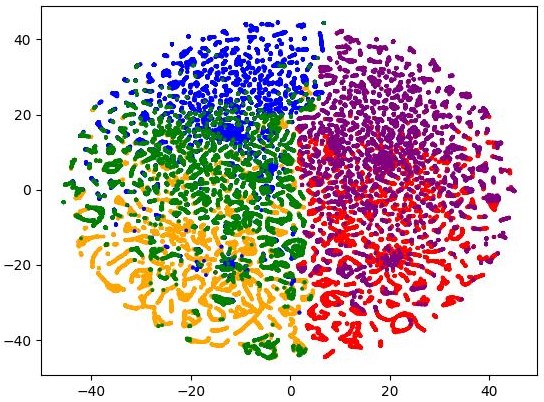}
	\end{minipage}
	\begin{minipage}[t]{0.19\linewidth}
		\includegraphics[width=1.0\linewidth]{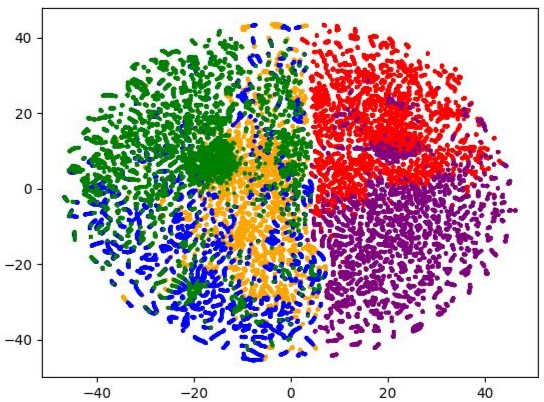}
	\end{minipage}
 	\begin{minipage}[t]{0.19\linewidth}
		\includegraphics[width=1.0\linewidth]{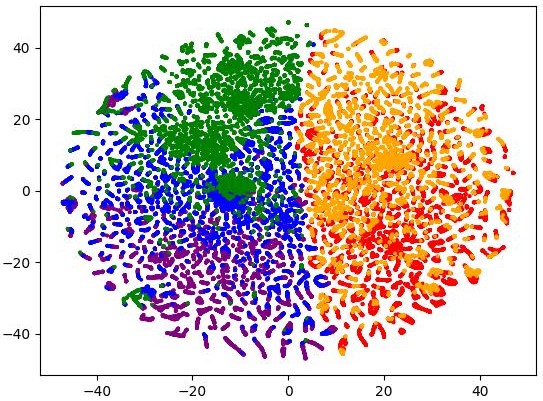}
	\end{minipage}
	\qquad

	\begin{minipage}[t]{0.19\linewidth}
		\includegraphics[width=1.0\linewidth]{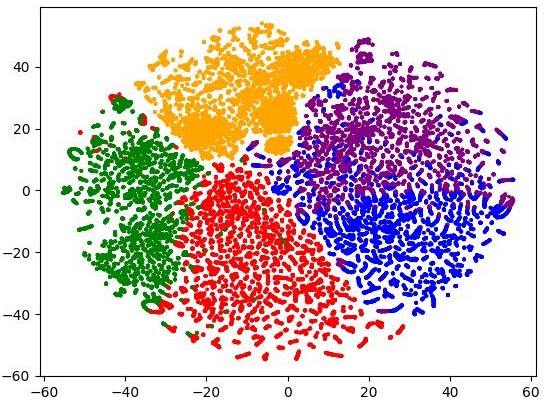}
	\end{minipage}
 	\begin{minipage}[t]{0.19\linewidth}
		\includegraphics[width=1.0\linewidth]{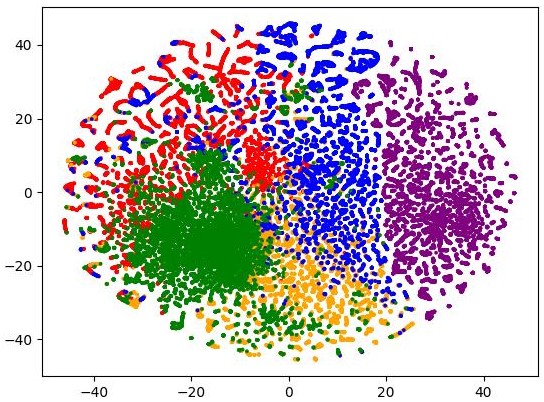}
	\end{minipage}
 	\begin{minipage}[t]{0.19\linewidth}
		\includegraphics[width=1.0\linewidth]{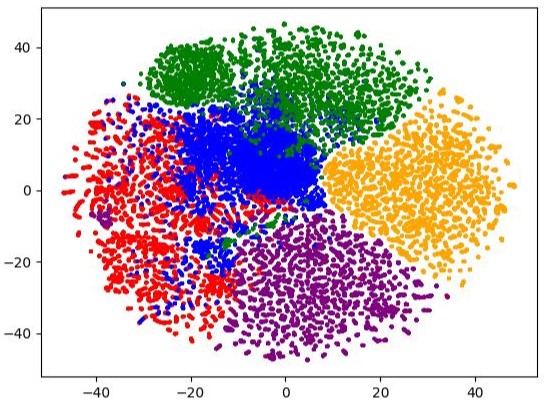}
	\end{minipage}
	\begin{minipage}[t]{0.19\linewidth}
		\includegraphics[width=1.0\linewidth]{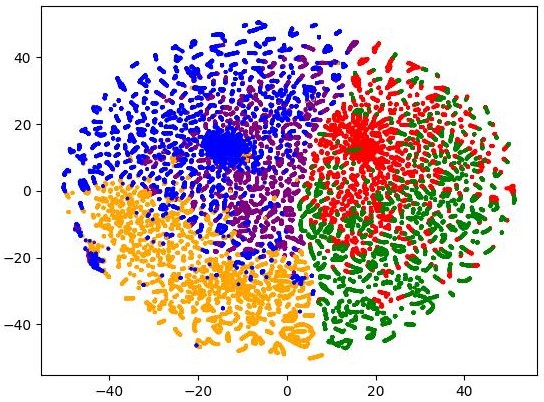}
	\end{minipage}
        \vspace{-3mm}
        \caption{T-SNE visualization of top 5 largest K-means/VAE-VQ clusters for 4 speakers,
        XM01, XM07, XM08, and XM11, (from left to right) with or without
phone purity guidance (lower and upper rows respectively)}\label{tsne}
\vspace{-6mm}
\end{figure*}
\subsection{Task Description}
The English UASpeech dysarthric corpus \cite{uaspeech2008} is an isolated word recognition task containing 103h speech recorded by 16 dysarthric speakers and 13 healthy control speakers.
For each speaker, the data is partitioned into three blocks B1, B2, and B3, with the same 155 common words and a different set of 100 uncommon words in each block.
The B1 and B3 data of all speakers serve as the training set while the data from B2 of the 16 dysarthric speakers is used as the test set.
Silence stripping performed via an HTK \cite{young2002htk} trained GMM-HMM system produces a 30.6h (99195 utt.) unaugmented training set and an 8.7h test set (26520 utt.).
All the dysarthric speech utterances in the unaugmented training set are used for phone purity analysis.
After further speed perturbation based data augmentation \cite{geng2020investigation}, a total of 130.1h speech data (399110 utt.) is used as the final training set for ASR system development. 
\vspace{-0.5em}

\subsection{Experiment Setup}
The hybrid LF-MMI factored time delay neural network (TDNN) systems \cite{peddinti15b_interspeech,povey16_interspeech} follow the Kaldi chain system setup, except that i-Vector are excluded.
The graphemic End-to-End (E2E) Conformer systems are implemented via ESPnet \cite{watanabe18_interspeech}.
A 512-dimensional embedding layer is inserted in the front of the E2E Conformer system for discrete token features.
\vspace{-6mm}


\subsection{Result Analysis}
Table \ref{tab:results} shows the performance comparison of hybrid TDNN and E2E Conformer systems constructed using different discrete token features. 
Several trends can be observed: 

\noindent\textbf{1)} Ppg. K-means discrete token features with varying numbers of clusters on TDNN systems consistently outperform the standard K-means by up to \textbf{0.99\% absolute (3.21\% relative)} WER reduction (Sys.9 vs Sys.8).
The Ppg. VAE-VQ method also consistently outperforms the non-Ppg. VAE-VQ token features on TDNN systems by up to \textbf{0.40\% absolute (1.31\% relative)} WER reduction (Sys.11 vs Sys.10).

\noindent\textbf{2)} The comparable experiment contrast confirms similar trends of 1) on Conformer systems. 
The Ppg. K-means token features consistently outperform the standard K-means token features by up to \textbf{1.77\% absolute (4.82\% relative)} WER reduction (Sys.19 vs Sys.18).
The VAE-VQ token features with phone purity guidance also consistently outperform VAE-VQ token features by up to 
\textbf{0.55\% absolute (1.57\% relative)} WER reduction (Sys. 21 vs Sys.20).

\noindent\textbf{3)} ASR performance improvements from phone purity guidance are further confirmed by those obtained on the phone purity metric in Table \ref{tab:purity}, and the more clearly defined cluster boundaries of Fig. \ref{tsne}, where the top 5 largest K-means/VAE-VQ clusters with or without phone purity guidance are shown in T-SNE plots for speakers XM01, XM07, XM08, and XM11.

\noindent\textbf{4)} System combination experiments were conducted to exploit the diversity amongst different token feature extraction schemes based on either K-means or VAE-VQ with or without phone purity guidance.
WER reductions of 6.43\% and 1.59\% absolute (20.85\% and 4.63\% relative, Sys. 31 vs Sys. 8 and Sys. 33 vs Sys. 22) respectively obtained for TDNN and Conformer systems over using standard unsupervised K-means or VAE-VQ tokens. 

\noindent\textbf{5)} Finally the most powerful form of system combination between the best performing TDNN (Sys. 34) and Conformer (Sys. 35) models using various types of token features produced the lowest WER of \textbf{23.25\%}, and outperformed the baseline TDNN and Conformer systems using standard K-means Hubert tokens (Sys. 36 Vs. Sys. 8, 22) by WER reductions of \textbf{7.59\% and 11.11\% absolute (24.61\% and 32.33\% relative)} respectively.
The performance of this final combined system is further contrasted against SOTA results of recent researches on the UAspeech task in table \ref{tab:comparison}. 

\begin{table}[!t]
    \vspace{-2mm}
    \caption{Phone purity comparison of discrete token features extracted by different methods with varying different codebook sizes.}
    \vspace{-2mm}
    \label{tab:purity}
    \centering
    \renewcommand\arraystretch{1.1}
    \begin{tabular}{c|c|c|c|c|c|c}
    \hline\hline
    \multirow{3}{*}{Methods} & \multirow{3}{*}{\shortstack{Code.\\Size}} & \multicolumn{5}{c}{Phone Purity \%} \\
    \cline{3-7}
    & & \multicolumn{4}{c|}{Intell. Subgroup} & \multirow{2}{*}{All}\\
    \cline{3-6}
    & & VL & L & M & H &  \\
    \hline\hline
    K-means & \multirow{4}{*}{100} & 32.37 & 34.14 & 36.62 & 42.02 & 36.29\\
    Ppg. K-means & & 32.70 & 34.76 & 37.13 & 42.67 & 36.84\\
    \cline{1-1}\cline{3-7}
    VAE-VQ & & 34.20 & 36.48 & 39.06 & 44.47 & 38.60\\
    Ppg. VAE-VQ & & 34.42 & 36.78 & 39.21 & 44.54 & 38.79\\
    \cline{1-7}
    K-means & \multirow{4}{*}{500} & 38.34 & 41.40 & 44.33 & 49.91 & 43.36 \\
    Ppg. K-means & & 38.66 & 41.93 & 44.87 & 50.34 & 43.81 \\
    \cline{1-1}\cline{3-7}
    VAE-VQ & & 39.24 & 41.87 & 44.62 & 49.75 & 43.72 \\
    Ppg. VAE-VQ & & 39.67 & 41.93 & 44.84 & 49.96 & 43.97\\
    \hline\hline
    \end{tabular}
    \vspace{-6mm}
\end{table}
\begin{table}[!ht]
    \vspace{-4mm}
    \centering
    \caption{A comparison between published systems on UASpeech and ours. Naming conventions follow the one adopted in Table \ref{tab:results}.}
    \vspace{-2mm}
    \label{tab:comparison}
    \renewcommand\tabcolsep{1pt}
    \resizebox{0.92\linewidth}{!}{
    \begin{tabular}{c|c|c|c}
    \hline
    \hline
    Performance (WER\%) of Systems Published on UASpeech& VL & L & All \\
    \hline
CUHK-2020 DNN + DA + LHUC-SAT \cite{geng2020investigation} & 62.44 & 27.55 & 26.37 \\
CUHK-2021 DNN + DCGAN + LHUC-SAT \cite{jin21_interspeech} & 61.42 & 27.37 & 25.89 \\
Brno Univ.-2022 Wav2vec2 + SAT (15 spkr) \cite{baskar22b_interspeech} & 57.72 & 22.46 & 22.83\\
FAU-2022 Cross-lingual XLRS + Conformer \cite{hernandez22_interspeech} & 62.00 & 28.60 & 26.10 \\
CUHK-2022 DNN + Data Aug. + SBE Adapt + LHUC-SAT \cite{mengzhe22SVD} & 59.30 & 26.25 & 25.05\\
CUHK-2023 Kaldi TDNN + VAE-GAN + LHUC-SAT \cite{jin23vae} & 57.31 & 28.53 & 27.78 \\
JHU-2023 DuTa-VC (Diffusion) + Conformer \cite{wang23qa_interspeech} & 63.70 & 27.70 & 27.90\\
CUHK-2023 TDNN + Wav2vec2.0 + Sys. Comb. \cite{hu2023exploring} & 53.12 & 25.03 & 22.56 \\
CUHK-2023 DNN + Wav2vec2.0 + Sys. Comb. + Sev. Adapt \cite{geng23b_interspeech} & 51.25 & 17.41 & 17.82 \\
CUHK- 2024 Wav2vec2.0/HuBERT + GAN Data Aug. + Sys. Comb. \cite{wang2024enhancing} & 46.47 & 16.76 & 16.53\\
\hline
\textbf{Ours, TDNN/Conformer + Discrete Token + Sys. Comb.} & \textbf{54.14} & \textbf{25.68} & \textbf{23.25} \\
\hline
\hline
\end{tabular}
}
\vspace{-3mm}
\end{table}

\section{conclusion}
This paper presents the first study on discrete token based dysarthric speech recognition and further proposes two phone-purity guided discrete token extraction methods to improve the quality of discrete tokens.
Experiments conducted on the UASpeech corpus suggest such supervised methods effectively improve the performance of discrete token based dysarthric speech recognition.
Future research will involve studying how to incorporate different supervision information into discrete token extraction.
\vspace{-1mm}
\section*{Acknowledgement}
This research is supported by Hong Kong RGC GRF grant No. 14200220, 14200021, 14200324, Innovation Technology Fund grant No. ITS/218/21, and NSFC Grant 62106255, Youth Innovation Promotion Association CAS Grant 2023119.

\newpage
\bibliographystyle{IEEEtran}
\bibliography{refs}

\begin{thebibliography}{10}
\providecommand{\url}[1]{#1}
\csname url@samestyle\endcsname
\providecommand{\newblock}{\relax}
\providecommand{\bibinfo}[2]{#2}
\providecommand{\BIBentrySTDinterwordspacing}{\spaceskip=0pt\relax}
\providecommand{\BIBentryALTinterwordstretchfactor}{4}
\providecommand{\BIBentryALTinterwordspacing}{\spaceskip=\fontdimen2\font plus
\BIBentryALTinterwordstretchfactor\fontdimen3\font minus \fontdimen4\font\relax}
\providecommand{\BIBforeignlanguage}[2]{{%
\expandafter\ifx\csname l@#1\endcsname\relax
\typeout{** WARNING: IEEEtran.bst: No hyphenation pattern has been}%
\typeout{** loaded for the language `#1'. Using the pattern for}%
\typeout{** the default language instead.}%
\else
\language=\csname l@#1\endcsname
\fi
#2}}
\providecommand{\BIBdecl}{\relax}
\BIBdecl

\bibitem{inproceedings}
H.~Christensen, M.~Aniol, P.~Bell, P.~Green, T.~Hain, and P.~Swietojanski, ``Combining in-domain and out-of-domain speech data for automatic recognition of disordered speech,'' in \emph{INTERSPEECH}, 2013.

\bibitem{xiong20source}
F.~Xiong, J.~Barker, Z.~Yue, and H.~Christensen, ``{Source Domain Data Selection for Improved Transfer Learning Targeting Dysarthric Speech Recognition},'' in \emph{ICASSP}, 2020.

\bibitem{liu21recent}
S.~Liu, M.~Geng, S.~Hu, X.~Xie, M.~Cui, J.~Yu, X.~Liu, and H.~Meng, ``{Recent Progress in the CUHK Dysarthric Speech Recognition System},'' \emph{IEEE/ACM TASLP}, 2021.

\bibitem{mengzhetaslp}
M.~Geng, X.~Xie, Z.~Ye, T.~Wang, G.~Li, S.~Hu, X.~Liu, and H.~Meng, ``Speaker adaptation using spectro-temporal deep features for dysarthric and elderly speech recognition,'' \emph{IEEE/ACM Transactions on Audio, Speech, and Language Processing}, 2022.

\bibitem{wang23y_interspeech}
T.~Wang, S.~Hu, J.~Deng, Z.~Jin, M.~Geng, Y.~Wang, H.~Meng, and X.~Liu, ``Hyper-parameter adaptation of conformer asr systems for elderly and dysarthric speech recognition,'' in \emph{INTERSPEECH}, 2023.

\bibitem{shujie23taslp}
S.~Hu, X.~Xie, M.~Geng, Z.~Jin, J.~Deng, G.~Li, Y.~Wang, M.~Cui, T.~Wang, H.~Meng, and X.~Liu, ``Self-supervised asr models and features for dysarthric and elderly speech recognition,'' \emph{IEEE/ACM Transactions on Audio, Speech, and Language Processing}, 2024.

\bibitem{wang2024enhancing}
H.~Wang, Z.~Jin, M.~Geng, S.~Hu, G.~Li, T.~Wang, H.~Xu, and X.~Liu, ``Enhancing pre-trained asr system fine-tuning for dysarthric speech recognition using adversarial data augmentation,'' in \emph{ICASSP}, 2024.

\bibitem{xiong2019phonetic}
F.~Xiong, J.~Barker, and H.~Christensen, ``{Phonetic Analysis of Dysarthric Speech Tempo and Applications to Robust Personalised Dysarthric Speech Recognition},'' in \emph{ICASSP}, 2019.

\bibitem{geng2020investigation}
M.~Geng, X.~Xie, S.~Liu \emph{et~al.}, ``{Investigation of Data Augmentation Techniques for Disordered Speech Recognition},'' \emph{INTERSPEECH}, 2020.

\bibitem{hu2023exploring}
S.~Hu, X.~Xie, Z.~Jin, M.~Geng, Y.~Wang, M.~Cui, J.~Deng, X.~Liu, and H.~Meng, ``{Exploring Self-supervised Pre-trained ASR Models for Dysarthric and Elderly Speech Recognition},'' in \emph{ICASSP}, 2023.

\bibitem{wang2023benefits}
P.~Wang and H.~Van~hamme, ``{Benefits of Pre-Trained Mono- and Cross-Lingual Speech Representations for Spoken Language Understanding of Dutch Dysarthric Speech},'' \emph{EURASIP J. Audio Speech Music Process.}, 2023.

\bibitem{jin21_interspeech}
Z.~Jin, M.~Geng, X.~Xie, J.~Yu, S.~Liu, X.~Liu, and H.~Meng, ``{Adversarial Data Augmentation for Disordered Speech Recognition},'' in \emph{INTERSPEECH}, 2021.

\bibitem{jin2024rl}
Z.~Jin, X.~Xie, T.~Wang, M.~Geng, J.~Deng, G.~Li, S.~Hu, and X.~Liu, ``Towards automatic data augmentation for disordered speech recognition,'' in \emph{ICASSP}, 2024.

\bibitem{baevski2019vq}
A.~Baevski, S.~Schneider, and M.~Auli, ``vq-wav2vec: Self-supervised learning of discrete speech representations,'' \emph{Proc. ICLR}, 2020.

\bibitem{baevski2020wav2vec}
A.~Baevski, Y.~Zhou, A.~Mohamed, and M.~Auli, ``wav2vec 2.0: A framework for self-supervised learning of speech representations,'' \emph{Advances in neural information processing systems}, 2020.

\bibitem{hsu2021hubert}
W.-N. Hsu, B.~Bolte, Y.-H.~H. Tsai, K.~Lakhotia, R.~Salakhutdinov, and A.~Mohamed, ``Hubert: Self-supervised speech representation learning by masked prediction of hidden units,'' \emph{IEEE/ACM Transactions on Audio, Speech, and Language Processing}, 2021.

\bibitem{chen2022wavlm}
S.~Chen, C.~Wang, Z.~Chen, Y.~Wu, S.~Liu, Z.~Chen, J.~Li, N.~Kanda, T.~Yoshioka, X.~Xiao, J.~Wu, L.~Zhou, S.~Ren, Y.~Qian, Y.~Qian, J.~Wu, M.~Zeng, X.~Yu, and F.~Wei, ``Wavlm: Large-scale self-supervised pre-training for full stack speech processing,'' \emph{IEEE Journal of Selected Topics in Signal Processing}, 2022.

\bibitem{chang23b_interspeech}
X.~Chang, B.~Yan, Y.~Fujita, T.~Maekaku, and S.~Watanabe, ``{Exploration of Efficient End-to-End ASR using Discretized Input from Self-Supervised Learning},'' in \emph{Proc. INTERSPEECH 2023}, 2023.

\bibitem{chang2024exploring}
X.~Chang, B.~Yan, K.~Choi, J.-W. Jung, Y.~Lu, S.~Maiti, R.~Sharma, J.~Shi, J.~Tian, S.~Watanabe \emph{et~al.}, ``Exploring speech recognition, translation, and understanding with discrete speech units: A comparative study,'' in \emph{ICASSP}, 2024.

\bibitem{yang2024towards}
Y.~Yang, F.~Shen, C.~Du, Z.~Ma, K.~Yu, D.~Povey, and X.~Chen, ``Towards universal speech discrete tokens: A case study for asr and tts,'' in \emph{ICASSP}, 2024.

\bibitem{sukhadia24_interspeech}
V.~N. Sukhadia and S.~A. Chowdhury, ``Children’s speech recognition through discrete token enhancement,'' in \emph{INTERSPEECH}, 2024.

\bibitem{du22b_interspeech}
C.~Du, Y.~Guo, X.~Chen, and K.~Yu, ``Vqtts: High-fidelity text-to-speech synthesis with self-supervised vq acoustic feature,'' in \emph{Interspeech}, 2022.

\bibitem{unicats}
C.~Du, Y.~Guo, F.~Shen, Z.~Liu, Z.~Liang, X.~Chen, S.~Wang, H.~Zhang, and K.~Yu, ``Unicats: A unified context-aware text-to-speech framework with contextual vq-diffusion and vocoding,'' \emph{Proceedings of the AAAI Conference on Artificial Intelligence}, 2024.

\bibitem{nguyen23_interspeech}
T.~A. Nguyen, W.-N. Hsu, A.~D'Avirro, B.~Shi, I.~Gat, M.~Fazel-Zarani, T.~Remez, J.~Copet, G.~Synnaeve, M.~Hassid, F.~Kreuk, Y.~Adi, and E.~Dupoux, ``Expresso: A benchmark and analysis of discrete expressive speech resynthesis,'' in \emph{INTERSPEECH}, 2023.

\bibitem{KENT2000273}
R.~D. Kent, J.~F. Kent, G.~Weismer, and J.~R. Duffy, ``What dysarthrias can tell us about the neural control of speech,'' \emph{Journal of Phonetics}, 2000.

\bibitem{uaspeech2008}
H.~Kim, M.~Hasegawa-Johnson, A.~Perlman, J.~Gunderson, K.~Watkin, and S.~Frame, ``{Dysarthric Speech Database for Universal Access Research},'' in \emph{INTERSPEECH}, 2008.

\bibitem{hernandez22_interspeech}
A.~Hernandez, P.~A. Pérez-Toro, E.~Noeth, J.~R. Orozco-Arroyave, A.~Maier, and S.~H. Yang, ``{Cross-lingual Self-Supervised Speech Representations for Improved Dysarthric Speech Recognition},'' in \emph{INTERSPEECH}, 2022.

\bibitem{zrvae}
Z.~Jin, X.~Xie, M.~Geng, T.~Wang, S.~Hu, J.~Deng, G.~Li, and X.~Liu, ``Adversarial data augmentation using vae-gan for disordered speech recognition,'' in \emph{ICASSP}, 2023.

\bibitem{AVSSL}
C.~Yu, X.~Su, and Z.~Qian, ``{Multi-Stage Audio-Visual Fusion for Dysarthric Speech Recognition With Pre-Trained Models},'' \emph{IEEE Transactions on Neural Systems and Rehabilitation Engineering}, 2023.

\bibitem{benefits}
P.~Wang and H.~Van~hamme, ``Benefits of pre-trained mono- and cross-lingual speech representations for spoken language understanding of dutch dysarthric speech,'' \emph{EURASIP J. Audio Speech Music Process.}, 2023.

\bibitem{pmlr-v202-radford23a}
\BIBentryALTinterwordspacing
A.~Radford, J.~W. Kim, T.~Xu, G.~Brockman, C.~Mcleavey, and I.~Sutskever, ``Robust speech recognition via large-scale weak supervision,'' in \emph{Proceedings of the 40th International Conference on Machine Learning}, ser. Proceedings of Machine Learning Research, A.~Krause, E.~Brunskill, K.~Cho, B.~Engelhardt, S.~Sabato, and J.~Scarlett, Eds., vol. 202.\hskip 1em plus 0.5em minus 0.4em\relax PMLR, 23--29 Jul 2023, pp. 28\,492--28\,518. [Online]. Available: \url{https://proceedings.mlr.press/v202/radford23a.html}
\BIBentrySTDinterwordspacing

\bibitem{kmeans++}
D.~Arthur and S.~Vassilvitskii, ``k-means++: the advantages of careful seeding,'' in \emph{Proceedings of the Eighteenth Annual ACM-SIAM Symposium on Discrete Algorithms}, 2007.

\bibitem{ruder2016overview}
S.~Ruder, ``An overview of gradient descent optimization algorithms,'' \emph{arXiv preprint arXiv:1609.04747}, 2016.

\bibitem{mapsswe}
L.~Gillick and S.~Cox, ``Some statistical issues in the comparison of speech recognition algorithms,'' in \emph{ICASSP}, 1989.

\bibitem{cui22_interspeech}
M.~Cui, J.~Deng, S.~Hu, X.~Xie, T.~Wang, S.~Hu, M.~Geng, B.~Xue, X.~Liu, and H.~Meng, ``Two-pass decoding and cross-adaptation based system combination of end-to-end conformer and hybrid tdnn asr systems,'' in \emph{Interspeech 2022}, 2022, pp. 3158--3162.

\bibitem{young2002htk}
S.~Young, G.~Evermann, M.~Gales, T.~Hain, D.~Kershaw, X.~Liu, G.~Moore, J.~Odell, D.~Ollason, D.~Povey \emph{et~al.}, ``{The HTK Book},'' \emph{Cambridge University Engineering Department}, 2002.

\bibitem{peddinti15b_interspeech}
V.~Peddinti, D.~Povey, and S.~Khudanpur, ``A time delay neural network architecture for efficient modeling of long temporal contexts,'' in \emph{Interspeech 2015}, 2015, pp. 3214--3218.

\bibitem{povey16_interspeech}
D.~Povey, V.~Peddinti, D.~Galvez, P.~Ghahremani, V.~Manohar, X.~Na, Y.~Wang, and S.~Khudanpur, ``Purely sequence-trained neural networks for asr based on lattice-free mmi,'' in \emph{Interspeech 2016}, 2016, pp. 2751--2755.

\bibitem{watanabe18_interspeech}
S.~Watanabe, T.~Hori, S.~Karita, T.~Hayashi, J.~Nishitoba, Y.~Unno, N.~{Enrique Yalta Soplin}, J.~Heymann, M.~Wiesner, N.~Chen, A.~Renduchintala, and T.~Ochiai, ``Espnet: End-to-end speech processing toolkit,'' in \emph{INTERSPEECH}, 2018.

\bibitem{baskar22b_interspeech}
M.~K. Baskar, T.~Herzig, D.~Nguyen, M.~Diez, T.~Polzehl, L.~Burget, and J.~Černocký, ``Speaker adaptation for wav2vec2 based dysarthric asr,'' in \emph{INTERSPEECH}, 2022.

\bibitem{mengzhe22SVD}
M.~Geng, X.~Xie, Z.~Ye, T.~Wang, G.~Li, S.~Hu, X.~Liu, and H.~Meng, ``Speaker adaptation using spectro-temporal deep features for dysarthric and elderly speech recognition,'' \emph{IEEE/ACM TASLP}, 2022.

\bibitem{jin23vae}
Z.~Jin, X.~Xie, M.~Geng, T.~Wang, S.~Hu, J.~Deng, G.~Li, and X.~Liu, ``{Adversarial Data Augmentation Using VAE-GAN for Disordered Speech Recognition},'' in \emph{ICASSP}, 2023.

\bibitem{wang23qa_interspeech}
H.~Wang, T.~Thebaud, J.~Villalba, M.~Sydnor, B.~Lammers, N.~Dehak, and L.~Moro-Velazquez, ``{DuTa-VC: A Duration-aware Typical-to-atypical Voice Conversion Approach with Diffusion Probabilistic Model},'' in \emph{INTERSPEECH}, 2023.

\bibitem{geng23b_interspeech}
M.~Geng, Z.~Jin, T.~Wang, S.~Hu, J.~Deng, M.~Cui, G.~Li, J.~Yu, X.~Xie, and X.~Liu, ``{Use of Speech Impairment Severity for Dysarthric Speech Recognition},'' in \emph{INTERSPEECH}, 2023.

\end{thebibliography}

\end{document}